\documentstyle[12pt,aasms4]{article}

\lefthead{Knezek, Sembach, and Gallagher}
\righthead{Dwarf Transition Galaxies}

\begin{document}

\title{Evolutionary Status of Dwarf ``Transition'' Galaxies}

\author{Patricia M. Knezek, Kenneth R. Sembach \altaffilmark{1}}
\affil{Department of Physics and Astronomy, The Johns Hopkins University, Baltimore, MD 21218-2695 \\ Electronic mail: pmk@pha.jhu.edu, sembach@pha.jhu.edu}
\and
\author{John S. Gallagher III \altaffilmark{1}}
\affil{Department of Astronomy, University of Wisconsin, Madison, WI 53706-1582 \\ Electronic mail: jsg@astro.wisc.edu}
\authoraddr{The Department of Physics and Astronomy, The Johns Hopkins University, Baltimore, MD 21218-2695} 
\altaffiltext{1}{Guest observer at the Michigan-Dartmouth-MIT Observatory}

\begin{abstract}

We present deep $B$-band, $R$-band, and H$\alpha$ imaging of 3 dwarf galaxies:
NGC~3377A, NGC~4286, and IC~3475.  Based on previous broadband imaging and 
\ion{H}{1} studies, these mixed-morphology galaxies have been proposed to be, 
respectively, a gas-rich low surface brightness Im dwarf, a nucleated
dwarf that has lost most of its gas and is in transition from Im to dS0,N, 
and 
the prototypical example of a gas-poor ``huge low surface brightness'' 
early-type galaxy (Sandage \& Hoffman 1991, Sandage \& Binggeli 1984). 
From the combination of our broadband and 
H$\alpha$ imaging with the published information on the neutral gas content of
these three galaxies, we find that (1) NGC~3377A is a dwarf spiral, similar
to those found by Schombert et al.\ (1995) and Matthews \& Gallagher (1997); 
(2) both NGC~3377A and NGC~4286 have 
comparable amounts of ongoing star formation, as indicated by their 
H$\alpha$ emission, while IC~3475 has no detected \ion{H}{2} regions to a
very low limit; (3) the {\it global} star formation rates are at least a 
factor of 20 below that of 30 Doradus for NGC~3377A and NGC~4286; 
(4) while the amount star formation is comparable, 
the {\it distribution} of star forming regions is very
different between NGC~3377A and NGC~4286, with H$\alpha$ emission scattered
over most of the optical face of NGC~3377A, and all contained within
the inner half of the optical disk of NGC~4286; (5) given their current 
star formation rates and gas contents, {\it both} NGC~3377A and NGC~4286 can
continue to form stars for more than a Hubble time; (6) both NGC~3377A and 
NGC~4286 have integrated total $B - R$ colors that are redder than the 
integrated total $B - R$ color for IC~3475, and thus {\it it is unlikely 
that either galaxy will
ever evolve into an IC~3475 counterpart}; and 
(7) {\it IC~3475 is too blue to be a dE}.  We thus conclude that we have not
identified potential precursors to galaxies such as IC~3475, and unless 
signifcant changes occur in the star formation rates, neither 
NGC~3377A nor NGC~4286 will evolve into a dwarf elliptical or dwarf 
spheroidal within a Hubble time.  
Furthermore,  
{\it optical morphology, even when coupled with the knowledge of the neutral
gas content of a dwarf galaxy, is not sufficient to determine its evolutionary
phase}.  
The evolutionary states of NGC~3377A and NGC~4286 are thus unclear and 
more complicated than might be inferred
from either previous broad-band imaging or \ion{H}{1} content alone. 

\end{abstract}


\section{Introduction}

An outstanding question in galaxy evolution is whether there is an
evolutionary connection between the various morphological classes of dwarf
galaxies. One view is that all dwarfs form in 
a similar manner, and the separation of dwarf irregulars (dSm/dIm) 
from early type dwarfs (dS0/dEs) takes place at a later evolutionary 
phase. For example, it has been suggested that dIm systems might evolve 
into dE or dS0 galaxies after they lose their gas (e.g., Lin \& 
Faber 1983, Thuan 1985). This might occur as a result of supernova 
driven winds during starbursts (Dekel \& Silk 1986, Babul \& Rees 1992, 
De Young \& Heckman 1994, Spaans \& Norman 1997), or as a result of 
an external interaction, such as ram pressure stripping (Kormendy 1985, 
Binggeli 1986) or galaxy collisions (Henriksen \& Byrd 1996, 
Moore et al. 1996, 1998). However, none of these mechanisms is able to reproduce 
the full range of differences between dIm and dE galaxies 
seen in the primary  
surface brightness-luminosity, luminosity-metallicity, and 
luminosity-rotation velocity correlations (and lack thereof in dEs) (see 
 Silk \&
Wyse 1993, Hunter \& Gallagher 1985, Bothun et al.\ 1986, Ferguson 
\& Binggeli 1994, Skillman \& Bender 1995, Walsh et al. 1997). At the 
present time we still do not understand the physical origins of the 
two main families of low luminosity galaxies.

Properties of `transition' dwarf galaxies, systems that show a mixture 
of early and late-type dwarf galaxy characteristics, therefore continue 
to be of special interest. 
Sandage \& Hoffman
(1991) proposed that the galaxies NGC~3377A, NGC~4286, and IC~3475 present an
example of the progression of gas-rich dwarfs (dSm/dIms) to gas-poor
dwarfs (dE/dS0s).  In a test to determine whether or not dIms could evolve 
into nucleated dwarf ellipticals, they chose two ``missing link'' galaxies
that had optical evidence from photographic plates 
of the mixed morphology of an 
underlying dE or dS0 with current star formation that was distributed 
globally over the optical surfaces of the galaxies.  

The first of these 
galaxies was NGC~3377A, known to be a gas-rich dwarf with many ``knots'' over
its face.  The second galaxy was NGC~4286, which also showed 
``knots'' over its face in a blue image, and a hint of spiral structure
in its ``outer envelope''.  Sandage \& Hoffman obtained \ion{H}{1} data for
NGC~4286 and found that the galaxy is gas-poor, although it 
still contains a moderate mass of \ion{H}{1}.  
Based on the combination of the optical morphology 
from photographic plates and the gas content from \ion{H}{1} data, they
concluded that NGC~4286 is indeed a galaxy in transition from Im to dS0,
while NGC~3377A has not yet been transformed. They proposed that 
NGC~3377A is similar to
the giant, very low surface brightness (VLSB) galaxy IC~3475.  IC~3475 is
a Virgo Cluster galaxy with little atomic gas, and a non-nucleated optical
morphology with statistical excess of marginally resolved 
embedded images over its face (Reaves 1956; Bothun et al.\ 1986).  Its optical
light distribution is best fit by an exponential light profile (Vigroux
et al.\ 1986), leading to the suggestion that it is a stripped dwarf
irregular.  To determine the validity of Sandage \& Hoffman's
proposition, we obtained deep broadband $B$ and $R$ CCD images, as well as
narrow-band H$\alpha$ images of these three galaxies 
to accurately determine their 
underlying stellar distribution and their current star formation rates
(SFRs).

In \S 2 we discuss the observations and reduction.  In \S 3 we
discuss
each galaxy individually and intercompare them.  We also discuss
the results in light of the picture proposed by Sandage \& Hoffman (1991).
\S 4 contains our conclusions.

\section{Observations and Data Reduction}

\subsection{Observations}

Direct imaging observations were made at Michigan-Dartmouth-MIT Observatory
on the 1.3 m McGraw--Hill 
telescope.  The data were taken on 10 - 12 April 1996 with 
a 1024${\rm x}$1024 thinned CCD.  The pixel scale is {$\sim$}0.50 {\arcsec}/pix
for this detector and telescope with the f/7.5 secondary, 
giving a field of view of 
{$\sim$}8{\farcm}5${\rm x}$8{\farcm}5.  Broadband $B$ and $R$ images were made
using the MDM Schombert filter set.  These filters closely 
approximate the wavelength coverage of the Kitt Peak National Observatory 
``Harris Set'' of broadband filters, although the transmission efficiencies 
are lower. Narrowband H$\alpha$ imaging was done using a set of extragalactic 
H$\alpha$ interference filters on loan from Deidre Hunter.  For NGC~3377A and
NGC~4286, a filter centered at {$\sim$}$\lambda$6590\AA\ was used.  For 
IC~3475, a filter centered at {$\sim$}$\lambda$6612\AA\ was used.  Both 
filters have a FWHM of {$\sim$}30\AA, and thus only include H$\alpha$
emission.  Typical integrations times were 
20-30 minutes in $B$, 10-15 minutes in $R$, and 40-60 minutes in H$\alpha$.  
Several Landolt standard fields (Landolt 1992) were observed each night in
the broadband filters, and the spectroscopic standard Feige 34 was observed
in $R$ and each H$\alpha$ filter.  The atmospheric extinction was determined
using the Landolt standards and found to be consistent with the standard 
Kitt Peak extinction coefficients.

\subsection{Data Reduction }

Initial data processing was done using the CCD reduction tasks in IRAF\footnotemark.  
\footnotetext{IRAF is distributed by the National Optical Astronomy
Observatories, which are operated by the Association of Universities
for Research in Astronomy, Inc., under contract to the National Science
Foundation.}
\noindent
The data were overscan-subtracted, and then bias subtracted 
using a median filtered bias from each night.  Primary flat fielding was
done using sky flats obtained each evening.  After applying the sky
flats typical variations were 1\% in $B$, 1.5\% in $R$, and 3\% in the two
H$\alpha$ filters.  These variations do not include the corners of the
images, where there was vignetting that was apparently due to
slight fluctuations in the filter wheel stopping point.  In no case does
the vignetting affect the portion of the image including the galaxies, and
the regions affected have been excluded from the remaining analysis.  When 
there were three or more individual exposures for a galaxy in one band, the
exposures in each band were aligned and then 
combined using a median filter with no rejection.  Any residual cosmic
rays were removed by hand.  For cases when there were only one or two exposures
of a galaxy in a band, cosmic rays were removed with the task 
{\it COSMICRAYS} in
IRAF, with any residual cosmic rays removed by hand.  If there were two
exposures, they were aligned and then combined by averaging with no
rejection.  Sky values for the final images were determined by taking the
average of $\sim$15 sky boxes around the images in regions selected to be
free from galaxy emission, stars, bad pixels, and vignetting.  The box size
was typically 15 pixels on a side.  The sky value was then subtracted from
the final images during the determination of the magnitudes and the 
radial surface brightness profiles of the galaxies.

In order to obtain the H$\alpha$ flux for each galaxy, the observations of
the spectrophotometric standard Feige 34 were used.  Feige 34 was observed
under photometric conditions in both the $R$ and H$\alpha$ filters.  
The transmission curve of each filter was convolved with the 
quantum efficiency of the CCD detector.  The resulting curve was then
multiplied by the atmospheric extinction curve.  (The KPNO extinction 
curve was used for this step, given that our broadband photometry implied
that the extinction during our run was consistent with it.)  
The spectral energy distributions of Feige 34 and BD$+$17{\arcdeg}4708 
were then convolved with the corrected efficiency curve, and the magnitude 
of BD$+$17{\arcdeg}4708 was set to 9.500 in each filter in order to place 
the magnitudes on the AB$_{\nu}$ system of Oke \& Gunn (1983).  The magnitude
of Feige 34 in each filter was then calculated relative to BD$+$17{\arcdeg}4708,
and the relationship between the magnitude in each filter and the AB$_{\nu}$ 
magnitude was found using the relationship of Windhorst et al.\ (1991):
\begin{displaymath}
AB_{\nu} = AB_{\nu}({\rm BD}+17{\arcdeg}4708) + m_{\lambda}(filter) - m_{\lambda}({\rm BD}+17{\arcdeg}4708)
\end{displaymath} 
where ``filter'' is either the $R$ filter or one of the H$\alpha$ filters, and
AB$_{\nu}$(BD$+$17{\arcdeg}4708) is calculated at the appropriate wavelength from
Oke \& Gunn (1983).  Using these magnitudes and the measured instrumental
magnitudes, a zero point was determined for each filter separately using the
relationship:
\begin{displaymath}
AB_{\nu} = (m_{\rm inst} - k_{\nu} * X_{\nu} ) + \zeta
\end{displaymath} 
The actual flux values could then be determined using the
relationship of Windhorst et al.\ (1991): 
\begin{displaymath}
AB(filter) = -2.5 {\rm log} F_{\nu}/d{\nu}~ ({\rm ergs~ cm^{-2}}~{\rm s}^{-1} ) + 48.594
\end{displaymath}

The total integrated H$\alpha$ magnitudes were measured at different 
points within each galaxy.   
For the two galaxies with positive detections of H$\alpha$ emission,
NGC~3377A and NGC~4286,
the radius was chosen such that it included all of the H$\alpha$ emission
visible in the image and avoided poorly subtracted stars.  (In the case of
NGC~3377A, one poorly subtracted star is included, which introduces an
uncertainty of $\sim 10$\% to the determined flux.)  For IC~3475, the reported
upper limit was calculated at a radius of 75\arcsec\ by taking three times
the reported counts within the aperture and converting that to an apparent
magnitude.  

\section{Discussion}

Our research was inspired by the suggestion by Sandage \& Hoffman
(1991) that the galaxies NGC~3377A, NGC~4286, and IC~3475 present 
examples of an evolutionary sequence running from typical gas-rich late-type
dwarfs (dSm/dIms) to gas-poor
dwarfs (dE/dS0s).
We obtained broadband $B$ and $R$ imaging, along with H$\alpha$ imaging
in order to compare the underlying stellar population, and ongoing star
formation (if any), with published neutral gas properties.

Table 1 lists the cataloged properties of the 3 galaxies in
this study.  Column 1 lists the galaxy name.  Columns 2 and 3 list the right
ascension and declination in J2000.0 coordinates.  Column 4 lists the 
heliocentric velocity in km s$^{-1}$.  Columns 5 and 6 list the major and minor
axes in arcminutes.  Column 7 lists the total $B$ magnitude.  Column 8 lists 
the \ion{H}{1} flux in Jy km s$^{-1}$.  Column 9 lists the morphological 
type.  All properties except the \ion{H}{1} flux were obtained from the
NASA/IPAC Extragalactic Database\footnotemark.  
\footnotetext{The NASA/IPAC Extragalactic Database (NED)
is operated by the Jet Propulsion Laboratory, California Institute
of Technology, under contract with the National Aeronautics and Space
Administration.}
\noindent
The \ion{H}{1} measurements are from Schneider et al.\ (1990)
for NGC~3377A, Sandage \& Hoffman (1991) for NGC~4286, and Huchtmeier
\& Richter (1986) for IC~3475.

%
%

\begin{table}
\dummytable\label{table1}
\end{table}

Table 2 presents the results of our photometry.  Column 1 lists the galaxy
name. Columns 2 and 3 list the total integrated magnitudes and magnitude errors 
of the galaxies
in the $B$ and $R$ bands, respectively.  These magnitudes were derived through
circular aperture photometry, and have not been corrected for 
contributions from foreground stars or for Galactic or internal 
extinction.  Column 
4 lists the total integrated magnitude of the H$\alpha$ emission after 
subtracting off the continuum.  In the case of IC~3475, the reported 
H$\alpha$ magnitude
is a 3$\sigma$ lower limit.  Because only one spectrophotometric standard was
observed, the errors on the H$\alpha$ magnitudes 
are uncertain.  However, the spectroscopically 
determined $R$ band flux for Feige 34
agrees to within $\sim$30\% of the broadband flux calculated using the 
Landolt (1992) standards, so we estimate that our H$\alpha$ measurements are
also accurate to $\sim$30\%.  We have included an additional 10\% uncertainty
for NGC~3377A due to the imperfectly subtracted star.  
Column 5 lists the total integrated $B - R$ 
color and its error at the last measured point for each galaxy, except for 
IC~3475.  For IC~3475 the light distribution is very flat and close to the
sky level in the outer regions, and the colors are dominated by possible
systematic errors in the sky level.  Thus we have chosen to quote the
color within 65\arcsec\ of the center, which is well within the region where 
the galaxy light still dominates.  The color is consistent with the color
at the last measured point.  The errors quoted for the magnitudes and colors
do not include possible systematic effects.  Column 6 lists the 
diameter of the galaxy at the $B_{25}$ mag arcsec$^{-2}$ isophote (corrected 
for Galactic extinction).  Column 7
lists the color excess due to Galactic extinction, $E(B-V)$.  This color excess
was derived using the standard dust-to-gas ratio given by Diplas \& Savage
(1994), 

\begin{displaymath}
E(B-V) = {4.93{\rm x}10^{21}\over{N(HI)}}
\end{displaymath}

\noindent
where $N(HI)$ was determined by integrating the \ion{H}{1} 21 cm emission
profile toward each galaxy as observed in the Leiden-Dwingeloo \ion{H}{1}
Survey (Hartmann \& Burton 1997).

%

\begin{table}
\dummytable\label{table2}
\end{table}

Table 3 lists the derived optical and neutral gas
properties of the 3 galaxies in this study.  Column 1 lists the 
galaxy name.  Column 2 lists the distances, which were adopted from
Tonry et al.\ (1997) for NGC~3377A and NGC~4286, and from
Garcia et al.\ (1996) for IC~3475.
Columns 3 and 4 list the absolute $B$ and $R$ 
magnitudes, respectively, based on magnitudes corrected for 
Galactic extinction.  No correction has been applied for internal extinction.  
Column 5 lists the H$\alpha$ luminosity.
All the optical data is based on our broadband 
and H$\alpha$ imaging. Column 6 lists the  
\ion{H}{1} masses, which 
were calculated using the fluxes from Table 1, the adopted
distances, and the relation 
M$_{HI} = 2.36{\rm x}10^{5} D^2 ~{\rm S}_{HI}$ M$_{\sun}$.  
Column 7 lists
the H$_2$ mass for NGC~4286, from Gerin \& Casoli (1994), scaled to our 
adopted distance.  Column 8 is 
a measure of the ratio of atomic 
gas to light, M$_{\rm HI}/{\rm L}_B$, where we
have adopted $M_B{\rm (Sun)} = +5.48$.  
Column 9 lists the current star formation rate  
calculated following the prescription of Kennicutt et al.\ (1994),
who rederived the relationship using the models of Schaller et al.\ (1993).
We have adopted a Salpeter Initial Mass Function.  The star
formation rate is then (see Kennicutt et al. 1994, Table 2):

\begin{displaymath}
{\rm SFR} = 1.26{\rm x}10^{-41} L({\rm H}\alpha)~ {\rm M}_{\sun}~ yr^{-1}
\end{displaymath}

\noindent
Column 10 indicates the amount of time that star formation can continue 
at the current level, 
estimated by dividing the current star formation rate by the 
available gas content. 
In this case, we have chosen to use the \ion{H}{1}
mass for the estimate, since the H$_2$ content has only been measured in
one galaxy, and the CO-H$_2$ conversion rate is uncertain.   This 
calculation assumes that the entire \ion{H}{1} mass is available for use
in star formation, and we have multiplied by a factor of 1.4 to include helium
and the heavy elements.

\begin{table}
\dummytable\label{table3}
\end{table}

\subsection{NGC~3377A}

NGC~3377A is located in the Leo I (M~96) 
group, and it is thought to be a companion to the E6 elliptical NGC~3377.  Its 
projected distance is only $\sim$22 kpc from NGC~3377.
It is gas-rich, M$_{HI}$/L$_B$ $\sim$ 0.30. In the $B$ image,
which can be seen in Figure 1a, its outer 
optical morphology is actually that of a late-type spiral, similar to the 
`extreme late-type' dwarf spirals.  
Figure 1b, which is a composite image of the $B$ band (represented as blue),
$R$ band (represented as red), and continuum subtracted H$\alpha$ (represented
by green), shows that it has a number of regions of active star formation spread
over almost the entire optical area of the galaxy, as well as regions of diffuse
H$\alpha$ emission.  The bright object near the ``center'' is either 
a foreground
star or a star cluster with colors comparable to the underlying stellar 
population.  There is no detected H$\alpha$ emission, so it is 
{\it not} an \ion{H}{2} region sinking to the center 
as suggested by Sandage \& Hoffman (1991).  

The radial surface brightness distribution of NGC~3377A was determined
within concentric circular annuli from a somewhat arbitrarily chosen
center.  The center was selected to be as close to the ``center'' of
the underlying central stellar distribution in $R$ as possible by
looking at the outer $R$ isophotes.  It lies approximately halfway 
between the bright stellar-like central region 
and a more diffuse light concentration of light (both free of
H$\alpha$ emission) approximately 
10\arcsec\ southwest (see Figure 1).  This diffuse light concentration may
be the dynamical center of the galaxy, or simply a large star cluster.  In
either case, it is slightly off-center in the $R$ band, and thus was not 
used as the center in this analysis.  The interactively determined
center was then used to
fit the magnitudes and surface brightness distributions in $B$, $R$, 
and H$\alpha$.  
The radial $B - R$ colors were determined both by subtracting the 
integrated $R$ magnitude from the integrated $B$ magnitude at the same 
radius, and by subtracting
the $R$ magnitude determined within an annulus from the $B$ 
magnitude within the same annulus.

Figure 2a shows the radial distributions for NGC~3377A.  The top panel 
contains the surface brightness distributions in $B$ (squares), $R$ 
(circles),
and H$\alpha$ (diamonds) as a function of radius.  
A typical error bar for the
$B$ and $R$ surface brightness distributions is shown in the upper right
portion of the top panel.  Errors in the H$\alpha$ distribution are dominated
by the uncertainty in the flux zero point, and are estimated to be $\sim$30\%.
Our calculated $B_T = 14.20\pm0.05$ 
agrees
well with de Vaucouleurs et al. (1991, hereafter RC3), who find 
$B_T = 14.22$.  The $B$ surface brightness distribution decreases by 
$\sim$ 1.5 magnitudes arcsec$^{-2}$ from the center to $\sim$ 90\arcsec.  
The dashed lines show simple linear fits to the data
using only points at $r > 60$\arcsec\ for $B$ and $R$.  This galaxy is fit
very well in the outer regions 
in both bands by the linear relationship, but there is evidence
of some excess light in the center.  The origin of this excess light is unclear
strictly from the optical images.
It is not apparent that there is any ``spheroidal'' component in this galaxy,
nor is there clear evidence for an optical bar, as is seen in IC~3475 (Vigroux
et al.\ 1986 and this paper).  This may be a case similar to those found by
Matthews \& Gallagher (1997) where there is a surface brightness ``step'' in
the disk.  Knezek (1993) also found that many objects in 
her sample of more massive
disk galaxies required the fit of a second ``disk'' component, independent 
of the presence of a well-fit bulge component, in order to fully model the
light distribution.  
The coefficients of the fit to the outer 60\arcsec\ 
are listed in Table 4. Column 1 lists the galaxy name.  Column 2 lists the
filter.  Column 3 lists the central surface brightness in mag arcsec$^{-2}$. 
Columns 3 and 4 list the disk scale length in arcsec and kpc, respectively.
This disk scale length was derived using the relationship (de Vaucouleurs 
1959):

\begin{displaymath}
\mu = \mu_0 + {r \over \alpha}.
\end{displaymath}

\noindent
The H$\alpha$ distribution, which was measured
in concentric annuli, decreases overall
as a function of radius, but with some structure.  
This non-linear distribution reflects the ``lumpy'' nature of the H$\alpha$
emission over the surface of the galaxy.  The global amount of H$\alpha$ 
emission is relatively low, L$_{H\alpha} \sim 2.6{\rm x}10^{38}$ erg s$^{-1}$,
where the H$\alpha$ luminosity of 30 Doradus is 
$\sim 5-10{\rm x}10^{39}$ erg s$^{-1}$ (Kennicutt \& Hodge 1986).  In fact, 
most of the quiescent dwarfs galaxies studied by van Zee et al.\ (1997a) 
have more H$\alpha$ luminosity (although there are some systems with 
comparable amounts).  In particular, for those galaxies of van Zee et al.\ 
(1997a) that are classified as dwarf spirals, most have $\sim$10 times more 
global H$\alpha$ emission.  Thus NGC~3377A appears to be a very quiescent 
dwarf spiral, despite its dense environment, and in this sense represents
a true transition object between low SFR dE and normal late-type dwarf
galaxies.

%
%

\begin{table}
\dummytable\label{table4}
\end{table}

The middle and bottom panels of Figure 2a show the integrated $B-R$ color and
annular $B-R$ color as a function of radius, respectively.  
NGC~3377A is blue over its entire optical disk, and becomes bluer with
increasing radius.  In the central region, its integrated $B - R$ 
is $\sim 1.05$, 
decreasing to $\sim 0.90$ in the outer regions of the disk.  The integrated 
and annular colors for this galaxy are very similar over the radial extent
for which we are able to reliaby measure the annular colors.  Both the 
integrated and annular colors show a fairly steep blueing ($\Delta(B - R) \sim
0.1 - 0.2)$ from the center out to $\sim$30\arcsec, then approximately 
level off, again indicating the possibility of two disk components.  
Furthermore, as can be seen in the bottom panel of Figure 2a, there is 
evidence that this second component is actually slightly {\it redder} than 
the inner disk component.  Reddening in outer disks was also seen in
some dwarf spirals in the sample of Matthews \& Gallagher (1997).  
For the disks 
of spiral galaxies, Bothun et al.\ (1985) found that the average $B - R$ 
color is 1.25.  Thus, NGC~3377A is bluer {\it at all points} than the typical
spiral galaxy disk.  The mean colors, however, are not
abnormally blue for dwarf and low surface brightness galaxies (Gallagher \& 
Hunter 1986, Knezek 1993,
Schombert et al.\ 1990, McGaugh \& Bothun 1994, de Blok et al.\ 1995, 
van Zee et al.\ 1996, van Zee et al.\ 1997a, van Zee et al.\ 1997b).  The 
colors are blue enough, however, that if most of the stellar population is
comparable in age to globular clusters around the Milky Way, it must be
very metal poor. Reed (1985) found that low metallicity Milky Way globular
clusters have mean colors, $< B - R > = 1.16$, {\it redder} than the colors
of NGC~3377A at all points. It seems unlikely, however, given the presence
of active star formation within its disk, that the colors are due only to
a very metal-poor old stellar population.

Despite its location in a dense environment, 
NGC~3377A resembles the dwarf spirals studied
by Matthews \& Gallagher (1997) and Schombert et al.\ (1995) in terms of
optical structure, \ion{H}{1} content, stellar luminosity, and colors.  
In fact, NGC~3377A
appears to be a fairly ``typical'' dwarf spiral in \ion{H}{1} 
gas mass, mass-to-light ratio, optical morphology, and colors.  It also
resembles the gas-rich quiescent dwarfs recently examined by van Zee
et al.\ (1997a) (many of which are also classified as dwarf spirals), 
although it is bluer.  (van Zee et al.\ find $B - R$ colors for their
sample of galaxies that are on average the same as those of Bothun et al.\
1985 for spiral disks.)  As can be seen in Table 3, NGC~3377A has a very low SFR
($\sim 0.0033$ M$_{\sun}$ ~yr$^{-1}$).  We have used a different prescription 
than van Zee et al.\ (1997a) to derive the SFR, but it should be noted that
our prescription implies a {\it higher} SFR than that of van Zee et al.  
Even so, our derived SFR is on the low end of those found by van Zee et al.,
and it is at least a factor of 10 lower than those of the galaxies 
classified as dwarf spirals in their sample.  The length of time that NGC~3377A
can continue to form stars at the current rate, however, is comparable, and
much longer than a Hubble time.  Thus, unless it is significantly affected by
its environment in the future, NGC~3377A is not going to evolve into a dE
anytime soon.  
Furthermore, it is unlikely to evolve into a galaxy like 
IC~3475, as its total integrated color within the detected broadband emission, 
$(B - R) \sim 1.15$ (corrected for Galactic extinction and with 
foreground stars masked) is already {\it redder} than IC~3475 
($(B - R) \sim 1.10$).

\subsection{NGC~4286}

NGC~4286 is a member of the Coma I group, at a projected distance of $\sim$36
kpc from the E1 galaxy NGC~4278, a \ion{H}{1}-rich elliptical.  
It is gas-poor, M$_{HI}$/L$_B$ $\sim$ 0.03, with a 
broadband optical mixed morphology between a late-type dwarf and a dwarf
elliptical or dwarf spheroidal.  
Figure 1c is the $B$ band image of this 
galaxy, which has been stretched to emphasize the faint spiral-like structure
noted by Sandage \& Hoffman (1991) 
in the inner part of the disk, where all the current star formation is 
occuring.  The outer regions of the galaxy, however, are quite smooth and
featureless, giving rise to the ``transition'' dS0 morphological
characteristics.  
Unlike NGC~3377A and IC~3475, this galaxy is
inclined.  This is confirmed not just from the optical appearance of the
outer isophotes (Figure 1c), but also from the fact that unlike NGC~3377A and
IC~3475, the \ion{H}{1} emission is double-horned (Sandage \& Hoffman 1991).  
Figure 1d, which is a composite image of the $B$ band (represented as blue),
$R$ band (represented as red), and continuum subtracted H$\alpha$ (represented
by green),
shows that the galaxy also has a number of active regions of star formation.  
The emission is more centrally concentrated than for NGC~3377A.  
Nearly all the 
detected H$\alpha$ emission for NGC~4286 lies within $\sim$40\arcsec,
which is $\sim$3 kpc.  The {\it total} extent of the H$\alpha$ emission of
NGC~3377A is comparable, lying within $\sim$75\arcsec, which corresponds
to $\sim$4 kpc.  However, since NGC~4286 is so much larger 
(D$(B_{25})$ $\sim$7.3 kpc for NGC~4286 versus $\sim$4.6 kpc for NGC~3377A), 
the relative concentration is very different.  This is also seen in the
upper panels of Figures
2a and 2b, where the radial distribution of H$\alpha$ is only slowly falling
with radius for NGC~3377A, but the radial distribution of H$\alpha$ drops
sharply for NGC~4286.
The bright object near the center {\it is} 
an \ion{H}{2} region that may have sunk to the center, as suggested 
by Sandage \& Hoffman (1991).  NGC~4286 has tenatively been 
detected in $^{12}$CO
(Gerin \& Casoli 1994), and the implied molecular-to-atomic gas ratio is
M$_{H2}$/M$_{HI}$ $\sim$ 0.23 (assuming a galactic conversion factor between
$^{12}$CO and H$_2$, Strong et al.\ 1988).  
This is comparable to ratios found in late-type disk
galaxies (Young \& Knezek 1989). Early-type disk galaxies have much higher 
ratios.  This ratio suggests that if NGC~4286 
is a galaxy in transition, it has lost much of its gas, not converted it
into molecular hydrogen.  Even if NGC~4286 is as metal-poor as local dwarf
irregulars such as IC~10 and NGC~6822, the increase in the 
molecular-to-atomic gas ratio is likely to be at most a factor of 3
(Wilson 1995). Then the {\it total} gas mass-to-light ratio is
M$_{gas}$/L$_{B}$ $\sim$ 0.05, and NGC~4286 is still a gas-poor galaxy.

Figure 2b shows
the radial surface brightness distribution of NGC~4286 along the major
axis. 
The position 
angle was taken from RC3. 
The top panel
contains the surface brightness distributions in $B$ (squares), $R$
(circles),
and H$\alpha$ (diamonds) as a function of radius.
The center was determined
within concentric circular annuli from the central peak of
the underlying central stellar distribution in $R$, which corresponds to
an \ion{H}{2} region. 
A typical error bar for the
$B$ and $R$ surface brightness distributions is shown in the upper right
portion of the top panel.   Errors in the H$\alpha$ distribution are dominated
by the uncertainty in the flux zero point, and are estimated to be $\sim$30\%.
The same center was then used to
fit the magnitudes and surface brightness distributions in $B$, $R$,
and H$\alpha$.  No $B_T$ was listed in RC3 to compare to our value of
$B_T = 13.71\pm0.05$.
The radial $B - R$ colors were determined both by subtracting the
integrated $R$ magnitude within a radius from the integrated $B$ magnitude
within the same radius (middle panel of Figure 2b), and by subtracting
the $R$ magnitude determined within an annulus from the $B$ 
magnitude within the same annulus (bottom panel of Figure 2b).
As a check, the radial
distributions were also calculated independently 
within concentric boxes of 10 by 10 pixels
along the major and minor axes.  
No significant differences in the shapes of the 
radial distributions of stellar light,
H$\alpha$ luminosity, or the color gradients were seen between the two axes.

As can be seen in the upper panel of Figure 2b, 
the radial surface brightness distribution for NGC~4286 deviates significantly
from a simple exponential disk.  Unlike NGC~3377A and IC~3475 (upper panels of
Figures 2a and 2c, respectively), which show only small deviations, 
NGC~4286 requires
a significant secondary component.  In fact, it appears 
that the secondary component dominates out to at least 60\arcsec, which is
0.6~D$(B_{25})$ (see below).  
To illustrate this point, we show the overlay of a simple linear regression 
analysis on radii $\gtrsim$~60\arcsec\ of the $B$ and $R$ images as a 
function of
radius.  
Again, this characteristic is
also found in some extreme late-type spirals (Matthews \& Gallagher 1997).  
The coefficients of
the fit to the outer 60\arcsec\ are listed in Table 3.  The 
H$\alpha$ surface brightness distribution increases from 5\arcsec\ to 
10\arcsec\ despite the fact that there is an \ion{H}{2} region at the center.
This may be due partly to internal extinction.
Outside of 10\arcsec\ the surface brightness decreases smoothly. It is
interesting to note that, unlike NGC~3377A, {\it all} of the H$\alpha$ 
emission detected in NGC~4286 lies within the inner ``secondary component'', 
and thus much of this inner light distribution, which is also bluer than the
outer regions of the galaxy (see below), may be due to the star formation.

From the middle and lower panels of Figures 2a and 2b, we can see that 
like NGC~3377A, NGC~4286 is blue over its entire optical disk. The
two galaxies
have similar annular $B - R$ colors in their inner 
10\arcsec\ ($(B - R)_{ann} \sim 1.05$).  
Except in the innermost point (which may 
be due simply to the fact that NGC~4286 center is an \ion{H}{2} region), the 
integrated and annular colors follow similar radial trends.  
However, 
while NGC~3377A becomes bluer with increasing radius, before levelling out
at $\sim$ 30\arcsec, NGC~4286 is nearly 
constant within the inner 15\arcsec\, and becomes slightly redder 
with
increasing radius, increasing to $(B - R)_{ann} \sim 1.15$ at 40\arcsec.  
The integrated $B - R$ colors also become redder until $\sim$40\arcsec, 
then remain approximately constant, $(B - R)_{int} \sim 1.10$,  
within D$(B_{25}) = 98$\arcsec.
There is no evidence
of any difference between the major and minor axes within the errors.  
The inner 40\arcsec\ is also the region where most of
the H$\alpha$ emission occurs, which is consistent with the blue colors.
Including a correction for internal extinction - most of which presumably
is in the center of the galaxy where the \ion{H}{2} regions are - would 
only enhance the difference in
colors between the inner and outer regions.  
The annular disk colors at greater than 40\arcsec\ (outside the regions of
active ongoing star formation) are comparable to the colors of IC~3475 at
a comparable radius, while the annular colors of NGC~3377A are bluer 
{\it over the entire optical face} (see the bottom panels of Figures 2a-c).  
This suggests that at least some regions of the disks of 
NGC~4286 and IC~3475 have older underlying 
stellar populations and/or significant chemical enrichment compared to
NGC~3377A.  Again, based on the study of low metallicity globular clusters 
by Reed (1985), if
the outer disk populations are old, they are very metal poor.  However, 
while the integrated colors of NGC~4286 truly level off to 
$(B - R)_{int} \sim 1.10$ out towards the edge of the detected optical disk, 
and perhaps even redden slightly ($(B - R)_{int} \sim 1.26$ at the outermost
measured point), 
the integrated colors of IC~3475 appear to get bluer with radius, 
and already have $(B - R)_{int} \sim 1.10$ within 65\arcsec.
Thus, NGC~4286 is {\it already too red} to fade into
a galaxy like IC~3475.  Furthermore, its SFR is comparable to that of
NGC~3377A, $\sim 0.0034$ M$_{\sun}$ ~yr$^{-1}$, again on the low end of
even the quiescent dwarfs studied by van Zee et al.\ (1997a). 
{\it Despite the fact that NGC~4286
is gas poor, assuming all of its gas is available to form stars, it can 
continue to form stars at its present level for more than a Hubble time}.

\subsection{IC~3475}

Garcia et al.\ (1996) derive a distance of 14.2 Mpc for IC~3475 and 
two ellipticals located nearby at about the same velocity, 
NGC~4168 (E2) and NGC~4473 (E6),
which would place them in front of the Virgo Cluster ($D = 17$ Mpc,
Kennicutt et al.\ 1995).  However, even if IC~3475 is not actually located near
the center of Virgo (projected distance $\sim$170 kpc from the Virgo Cluster
core, at the distance of M~100), it appears to be
in a dense
environment with nearby elliptical galaxies, similar to NGC~3377A and NGC~4286. 
Like NGC~4286, it is also gas-poor, M$_{HI}$/L$_B$ $<$ 0.09, with a fairly
smooth, diffuse morphology characteristic of a dwarf elliptical or a dS0.  
Since it is not significantly centrally concentrated, its 
classification is closer to dS0 than dE (c.f.\ Ferguson \& Binggeli 1994).
Figure 1e is the $B$ band image of this
galaxy, which has been stretched to emphasize the faint bar-like structure in
the central region, which was noted by Vigroux et al.\ (1986).
Figure 1f, which is a composite image of the $B$ band (represented as blue),
$R$ band (represented as red), and continuum subtracted H$\alpha$ (represented
by green), shows that 
there
is no evidence of any H$\alpha$ emission at a 3$\sigma$ flux level of
$f_{\nu} = 5.15{\rm x}10^{-14}$ ergs cm$^{-2}$ s$^{-1}$ within a radius of
75\arcsec\ from the ``center''.  Also, as found by Vigroux et al.\ (1986), 
none of the ``bumps'' have
anomalously blue colors.  Furthermore, if IC~3475 had any \ion{H}{2} regions
with emission comparable to the fainter \ion{H}{2} regions in 
NGC~3377A or NGC~4286, they would have been detectable with our data.  
The 3$\sigma$ detection
limit for a potential 
\ion{H}{2} region in IC~3475, scaled in size by the difference
in distances, is almost a factor of 2 below the flux detected in a comparable 
faint \ion{H}{2} region in NGC~3377A.  Thus, 
this galaxy does not appear to have any significant ongoing
star formation, nor does there appear to have been any star formation in 
the recent past.  

Figure 2c shows the radial distributions for IC~3475.  The top panel
contains the surface brightness distributions in $B$ (squares), $R$
(circles),
and H$\alpha$ (diamonds) as a function of radius.
A typical error bar for the
$B$ and $R$ surface brightness distributions is shown in the upper right
portion of the top panel.  Errors in the H$\alpha$ distribution are dominated
by the uncertainty in the flux zero point, and are estimated to be $\sim$30\%.
The radial surface brightness distribution of IC~3475 was determined
within concentric circular annuli from a somewhat aribitrarily chosen
center.  The center was selected to be as close to the ``center'' of
the underlying central stellar distribution in $R$ as possible by
looking at the outer $R$ isophotes and the central bar-like structure.  
The same center was then used to
fit the magnitudes and surface brightness distributions in $B$, $R$,
and H$\alpha$.
The radial $B - R$ colors were determined 
both by subtracting the
integrated $R$ magnitude within a radius from the integrated $B$ magnitude
within the same radius (middle panel of Figure 2c), and 
the $R$ magnitude determined within an annulus from the $B$ 
magnitude within the same annulus (bottom panel of Figure 2c).  The error
bars probably {\it underestimate} the error in the colors outside of
a radius of $\sim$ 120\arcsec.  In the $R$-band, where the sky is much
brighter, we find that outside $\sim$ 120\arcsec\ a change of 1 count/pixel
in the sky 
(which is the standard deviation of the mean of the sky in this image) 
can produce a $\sim$ 0.1 mag variation in the total magnitude.
Thus, we have chosen to be conservative and will only use colors within
65\arcsec, where the variation is seen to be $\lesssim 0.05$ mag.

Our calculated $B$ magnitude, $B_{T} = 13.95\pm0.05$, is a magnitude brighter
than that reported by Vigroux et al.\ (1986, $B_{T} = 14.7$), 
but close to the RC3
value, $B_{T} = 13.82$.  However, our determined {\it central} 
surface brightness, $\mu_0(B) = 23.65$, agrees well with the value 
measured by Vigroux et al., $\mu_0(B) = 23.60$.  
Vigroux et al.\ (1986) pointed out that IC~3475
nearly filled their field of view, and that their sky background may have
been incorrectly subtracted.  This also leads to significant differences
in the radial surface brightness distributions between our study and theirs.
In contrast to Vigroux et al.\ (1986),
who find a decrease of almost 3.5 magnitudes arcsec$^{-2}$ between the
center and their last measured point at $\sqrt{ab} \sim 70$\arcsec, we see
a decrease of $\lesssim$ 1.5 magnitudes arcsec$^{-2}$ over the same
radius, again consistent with a sky subtraction difficulty in the previous
data.  
This difference is unlikely to be due to the isophotal photometry methods
used (circular apertures in this study, elliptical apertures in the 
Vigroux et al.\ (1986) study)
since IC~3475 has a small ellipticity (typically $e \lesssim 0.25$,
Vigroux et al.\ (1986)). 
As can be seen in the upper panel of 
Figure 2c, IC~3475 has an exponential surface brightness
distribution in both $B$ and $R$, but with a small excess in the
inner regions.  This
deviation from linearity may well be due to the presence of the bar described by
Vigroux et al.\ (1986).  The coefficients of
the fit to the outer 60\arcsec\ are listed in Table 3.  The excess light
component appears to be similar to that seen in NGC~3377A, and much less
extreme than the deviation seen in NGC~4286 (see the upper panels of Figures
2a and 2b).

The middle and lower panels of Figure 2c indicate that both
the integrated and annular 
$B - R$ colors of IC~3475 are only slightly bluer than the disk of a 
typical spiral
galaxy in its inner regions ($(B - R) \sim 1.15 - 1.2$, $r \lesssim 30$\arcsec). 
The integrated colors decrease very slowly 
to a $(B - R)_{int} \sim 1.10$ out to a radius of $r \sim 65$\arcsec, while the
annular colors have dropped to $(B - R)_{ann} \sim 1.0$ within the same
radius.
Given the lack of gas or ongoing star 
formation, along with the $B - R$ colors, it appears that IC~3475 
has completed its last episode of star formation and is now fading.
As noted in the section above, its annular 
colors are comparable to the outer disk of NGC~4286, and only in the 
outermost areas 
do they resemble those of NGC~3377A.  The inner regions 
($\sim$40\arcsec) have colors which are comparable to those of the low
metallicity globular clusters (Reed 1985), while the outer regions are 
bluer and imply that if the stellar population is old and fading, it must
be quite metal-poor.   

The suggestion that IC~3475 is not a ``typical'' dE galaxy is further
strengthened by its comparison to the sample of dEs studied by Secker
et al.\ (1997) in the Coma Cluster.  They found a very tight correlation
between the total $R$ magnitude of the Coma dEs and their $B - R$ colors
(see their Figure 4).  Even without correcting for Galactic or internal
extinction, IC~3475 has an $R$ magnitude of $12.83\pm0.05$, which 
corresponds to an apparent magnitude of $R \sim 16.89$ at the distance of
Coma, and a 
$B - R$ color of $1.10\pm0.04$ (see Table 2).  Assuming that the $B - R$
color is distance independent, and using the relationship 
they give in their equation 5 between $R$ magnitude and $B - R$ color, 
IC~3475 is $\sim$0.4 magnitudes {\it too blue} to be a dE.

\subsection{Evolutionary Progression and Environmental Effects}

Our data are only 
partially consistent with the hypothesis proposed by Sandage \& 
Hoffman (1991) that NGC~3377A, NGC~4286, and IC~3475 represent dwarf 
galaxies in various stages of transition from late to early-types.  
NGC~3377A appears to be a comparatively normal extreme late-type spiral, while
in NGC~4286 the star formation has become centrally concentrated within the stellar
disk.  As a result the {\it outer} regions of NGC~4286 now resemble those of a
dS0 galaxy, but, its integrated color is already {\it too red} for it to 
fade into a galaxy resembling IC~3475 ($B - R \sim 1.10$).
IC~3475 appears to be a dE/dS0 in which no recent star formation has
occurred.  Its position on a $(B - R)$-$R$ color-magnitude diagram 
(Secker et al.\ 1997) is inconsistent with classification as a ``typical''
dE. 
What is not clear is whether these galaxies are currently 
experiencing abnormal evolution towards earlier structural types.  The gas
resevoirs in both NGC~3377A and NGC~4286 are adequate to sustain their current
characteristics for long times.  We see no evidence for rapid evolutionary
transitions of NGC~4286 or NGC~3377A.  If 
NGC~4286 is a dwarf galaxy ``in transition'', as is indicated by its gas-poor
nature and its regular outer isophotes, it will {\it never} fade into a dwarf
``stellar fossil'' galaxy like IC~3475, although the possibility exists in terms of $B-R$ colors
for NGC~4286 to have the {\it normal} colors of a dE or dS0 galaxy if star
formation were to cease.

The effects of environment on the evolution of small galaxies remains 
relatively poorly understood.  Researchers
have proposed that the evolution of dwarf galaxies may be driven by internal
mass loss due to stellar winds (Sandage 1965; Searle \& Zinn 1978; Faber
\& Lin 1983; Wirth \& Gallagher 1984; Matteucci \& Chiosi 1983; Matteucci
\& Tosi 1985; Pilyugin 1992, 1993; Marconi et al.\ 1994) 
or by the loss of gas due to
stripping in a group or cluster environment (Lin \& Faber 1983).  However,
the picture is not a simple one because the gas distribution, and 
presumably the subsequent star formation history and stellar light
distribution, can be driven by the {\it type} of environmental interactions,
and their significance
as compared with internal processes.

Galaxies, particularly dwarf galaxies, in groups and clusters are
subject to a variety of environmental influences. The presence of hot
intergalactic gas could lead to stripping, especially from the low density
outer regions of dwarfs (e.g., Gallagher \& Hunter 1989), but the intra-group
gas could also cool and accrete, thereby rejuvenating previously moribund
dwarfs (Silk et al.\ 1987). Gravitational interactions between
dwarfs and giants will dynamically heat the stars in the dwarfs and can
also raise substantial tides (McGlynn \& Borne 1991, Gruendl et al. 1993,
Valluri 1993, Oh et al.\ 1995, Moore et al.\ 1998).
When interactions are strong or occur repeatedly, profound changes in
the internal mass distributions of dwarfs can occur, including
complete disruption (e.g., S\'eguin \& Dupraz 1996). Close passages
near gas-rich giants
can also lead to gas stripping from a smaller galaxy (Sofue 1994).  Other,
less intense collisions can redistribute gas, either to be lost into
tidal streams (e.g., Gardiner \& Noguchi 1996) or to lose angular momentum
and drop into the center of the dwarf (e.g., Noguchi 1988).

A variety of intra-group interactions are therefore capable of
converting a dwarf galaxy from a gas-rich to a gas-poor state.
However, relatively strong interactions are required to change a
rotationally supported stellar disk of a Magellanic or extreme
late-type spiral galaxy, which has a ratio of stellar rotation
to dispersion velocity that is considerably larger than 1, 
into a non-rotating dE system where this ratio is $<$1 (Bender \&
Nieto 1990, Moore et al. 1998). Transition dwarfs that are not
directly involved in strong interactions are therefore likely to
be evolving towards gas-poor disk galaxies, or dS0 systems.

Modeling of galaxy interactions indicates that whether gas is 
stripped from a system or driven to the center of the galaxy
can be a sensitive function of the location of the
galaxy relative to other systems.  NGC~3377A is a dwarf companion to the
E6 galaxy NGC~3377, at a projected distance of only $\sim$22 kpc (Sandage \&
Hoffman 1991).  NGC~4286 is also a dwarf companion galaxy, in this
case to the giant \ion{H}{1}-rich elliptical NGC~4278, with a projected linear 
separation of only $\sim$36 kpc (Sandage \& Hoffman 1991).  
The possibility therefore exists that the unusual structure of NGC~4286
could be associated with a past interaction between it and its giant
elliptical neighbor (e.g.\ DuPrie \& Schneider 1996).  IC~3475 is 
is only $\sim$170 kpc from the projected center of the Virgo Cluster (Vigroux
et al.\ 1986), 
apparently placing it well within the cluster core.  Even if it is actually
in front of Virgo, as indicated by the recent distant estimates of Garcia
et al.\ (1996), it is still apparently a member of a group with massive
ellipticals.  Thus, the 
evolution of all three
galaxies may have been affected by their local environments.

The evolutionary history of these galaxies remains obscure.  
Additional information may be
provided by the kinematics
of the gas and the metallicities of the galaxies.  
By simultaneously determining the chemical enrichment of the ionized gas
and studying the gas kinematics, we may be able to resolve these issues.
Under the auspices of the ``transition from dIm to dE/dS0'' theory, one would
expect that high atomic gas-to-light ratio galaxies in a pre-starburst
phase would have lower abundances and relatively undisturbed gas kinematics,
while galaxies that have just begun the starburst process would also have
lower abundances and kinematical signatures of 
active star formation and supernovae-driven winds.  
Dwarfs
that have progressed further along the transition phase may show more
chemical enrichment, lower atomic gas-to-light ratios, and a kinematical
signature of supernovae-driven winds.  Galaxies that have entered the
postburst phase should also be chemically enriched, but no longer show
signs of ongoing massive star formation.  Abundance and kinematical
information are necessary to determine if this scenario holds under scrutiny
and whether it depends upon environmental considerations.
If the kinematics and
chemical abundances of NGC~3377A are consistent with the idea that it is in an
earlier, less enriched, stage of development compared to NGC~4286,
and the kinematics of NGC~4286 are consistent with 
gas loss through stripping or supernova driven winds, then we will
have a indication that this hypothesis is valid.  The relation of IC~3475 
to other dwarf galaxies remains unclear, however.  
In order to examine
this, we plan to obtain high resolution spectroscopy of the galaxies
studied here.

\section{Conclusions}

We have examined the galaxies NGC~3377A, NGC~4286, and IC~3475 in $B$, $R$, 
and H$\alpha$.  
NGC~3377A is actively forming stars over most of its optical area, but 
at a very low level.   It is quite
gas-rich, and appears to be a dwarf spiral. 
If all of its gas is available to form stars, it
can continue to maintain its current star formation level for much
longer than a Hubble time, and this, along with its $B - R$ colors, argue
against it evolving into an IC~3475-type galaxy
in the near future.  
It may be a precursor to a galaxy resembling
NGC~4286, but
whether it can {\it physically} evolve into such a galaxy 
later in its evolution is unclear without further observations.  
NGC~4286 is a dwarf that
may have lost most of its gas very recently, perhaps through stripping, or 
through supernovae-driven winds as it underwent an environmentally-induced
episode of star
formation.  All of its current star formation is located in the inner half 
of the optical disk.  
At its current low SFR, however, like NGC~3377A, 
it can continue to form stars for
more than a Hubble time.  Furthermore, its $B - R$ color is also inconsistent
with the suggestion that it may
evolve into a galaxy resembling IC~3475.
IC~3475 appears to be a quiescent, gas-poor, low
surface brightness galaxy, with no signs of ongoing or recent star formation
activity.  However, it is {\it too blue} to be a dE, and no obvious local
counterpart to its precursor has yet been identified.  
Thus the evolutionary progression of these dwarf galaxies
remains uncertain, and we find that 
using optical morphology, even when combined with some
knowledge of the neutral gas content of a dwarf galaxy, is {\it not} 
sufficient to determine a galaxy's evolutionary phase.

Other studies of ``transition'' dwarf galaxies reach similar 
conclusions that knowledge of optical morphology is not 
sufficient to constrain the evolutionary phase of dwarfs.  For example, 
Sandage \& Fomalont (1993) suggested based on red and yellow images, that
the galaxy ESO~359-029 is a ``transition dwarf'' very similar to 
NGC~4286.  Based on the photometry of Lauberts \& Valentijn (1989), they
derive $(B-R)_T = 0.63$, which is quite blue.  
Their follow-up \ion{H}{1} observations, however, indicated that 
this galaxy is {\it not} gas poor, like NGC~4286, for which they derive
M$_{HI}$/L$_{B} =$ 0.21. This value is much
closer to NGC~3377A than NGC~4286.  Sandage \& Fomalont then concluded that
ESO~359-029 is a ``transition dwarf'' {\it between} the stages of NGC~3377A
and NGC~4286.  In fact, recent CCD observations of 
ESO~359-029 show this galaxy to have the 
morphology of an extreme late-type Sd spiral (Matthews \& Gallagher 
1997).  Furthermore, Matthews \& Gallagher derive moderately red optical 
colors, $(B-V)_T = 0.62$, at odds with the blue colors found by Sandage \&
Fomalont.  Also, despite moderately red colors, \ion{H}{2} regions and 
diffuse emission from ionized gas are also seen in  
a long slit spectrum (Matthews 1998), while high signal-to-noise 
pencil beam \ion{H}{1} observations show a normal line profile and 
M$_{HI}$/L$_V =$0.43 (Matthews et al.\ 1998). This system is
neither gas-poor, like NGC~4286, nor a drop out from star-forming activity, 
like NGC~3377A.  Given the conflicting results for the optical colors, it
also remains unclear whether this galaxy, even if it lies between the
evolutionary stages of NGC~3377A and NGC~4286, as suggested by Sandage \&
Fomalont, could ever evolve into a galaxy like IC~3475.

Obviously, in order to truly begin to deconvolve the 
evolution of dwarf galaxies, 
a statistically significant sample needs to be studied.  This sample needs
to be selected to distinguish between environmental and internal effects.
Furthermore, 
information about the kinematics
of the gas and the metallicities of the galaxies 
is also needed to disentangle
the different possible evolutionary paths.  The study presented here
represents the first step in an ongoing project to accomplish these goals.

\acknowledgements

The H$\alpha$ filter set used for this research 
was kindly loaned to us by Deidre
Hunter.  Rosa Gonzalez provided us with the scripts for the spectrophotometric
data reduction, Ricky Patterson offered many helpful comments on the
surface photometry, and Gerhardt Meurer provided helpful discussions.  
In addition, we would like to thank
the staff at the Michigan-Dartmouth-MIT Observatory for their
assistance and support in this project.  PMK and KRS acknowledge support
from NASA Long Term Space Astrophysics grant NAG5-3485.  JSG wishes to thank
the National Science Foundation for partial support of his research on dwarf
galaxies through grant NSF AST-9803018.

\vfill
\eject

\figurenum{1a}
\figcaption[knezek.fig1a.ps]{ The image shows a greyscale
4.2$^{\prime}$x4.2$^{\prime}$
$B$-band image of NGC~3377A. A bar representing 15\arcsec\ is in the lower 
right corner.  North and east are indicated. \label{fig1a}}

\figurenum{1b}
\figcaption[knezek.fig1b.ps]{ The image is a 4.2$^{\prime}$x4.2$^{\prime}$ three-color
image of NGC~3377A.  Blue represents $B$-band, red represents $R$-band,
and green represents the continuum-subtracted H$\alpha$ emission.  Colors
have been determined to maximize contrast.
North and east are indicated. \label{fig1b}}

\figurenum{1c}
\figcaption[knezek.fig1c.ps]{ The image shows a greyscale 
4.2$^{\prime}$x4.2$^{\prime}$
$B$-band image of NGC~4286. A bar representing 15\arcsec\ is in the lower 
right corner.  North and east are indicated.  \label{fig1c}}

\figurenum{1d}
\figcaption[knezek.fig1d.ps]{ The image is a 4.2$^{\prime}$x4.2$^{\prime}$ three-color
image of NGC~4286.  Blue represents $B$-band, red represents $R$-band,
and green represents the continuum-subtracted H$\alpha$ emission.  Colors
have been determined to maximize contrast.
North and east are indicated.  \label{fig1d}}

\figurenum{1e}
\figcaption[knezek.fig1e.ps]{ The image shows a greyscale 
4.2$^{\prime}$x4.2$^{\prime}$
$B$-band image of IC~3475. A bar representing 15\arcsec\ is in the lower 
right corner.  North and east are indicated.  \label{fig1e}}

\figurenum{1f}
\figcaption[knezek.fig1f.ps]{ The image is a 4.2$^{\prime}$x4.2$^{\prime}$ 
three-color
image of IC~3475.  Blue represents $B$-band, red represents $R$-band,
and green represents the continuum-subtracted H$\alpha$ emission.  Colors
have been determined to maximize contrast.
North and east are indicated.  \label{fig1f}}

\figurenum{2a}
\figcaption[knezek.fig2a.ps]{ The upper panel 
shows the $B$ (filled squares), $R$
(filled circles), and H$\alpha$ (filled triangles) 
surface brightnesses as a function of radius for NGC~3377A.  
The dashed lines are the best
fit to a line using only the data at $r > 60$\arcsec\ in $B$ and $R$.
The middle panel shows the integrated 
$B - R$ color as a function
of radius within 5\arcsec\ annuli.
The bottom panel shows the annular 
$B - R$ color as a function
of radius within 5\arcsec\ annuli.}

\figurenum{2b}
\figcaption[knezek.figbd.ps]{ The upper panel 
shows the $B$ (filled squares), $R$
(filled circles), and H$\alpha$ (filled triangles) 
surface brightnesses as a function of radius for NGC~4286 along the major axis.
The dashed lines are the best
fit to a line using only the data at $r > 60$\arcsec\ in $B$ and $R$.
The middle panel shows the
integrated $B - R$ color as a function
of radius within 5\arcsec\ annuli. 
The bottom panel shows the 
annular $B - R$ color as a function
of radius within 5\arcsec\ annuli.}

\figurenum{2c}
\figcaption[knezek.fig2c.ps]{ The upper panel
shows the $B$ (filled squares), $R$
(filled circles), and H$\alpha$ (filled triangles) 
surface brightnesses as a function of radius for IC~3475.  The H$\alpha$ 
surface brightnesses are 3~$\sigma$ upper limits.
The dashed lines are the best
fit to a line using only the data at $r > 60$\arcsec\ in $B$ and $R$.
The middle panel shows the integrated 
$B - R$ color as a function
of radius within 5\arcsec\ annuli.
The bottom panel shows the annular 
$B - R$ color as a function
of radius within 5\arcsec\ annuli.}

\end{document}